\documentclass[conference]{IEEEtran}
\IEEEoverridecommandlockouts
\usepackage{cite}
\usepackage{amsmath,amssymb,amsfonts}
\usepackage{algorithmic}
\usepackage{graphicx}
\usepackage{textcomp}
\usepackage{lipsum}
\usepackage{xcolor}
\usepackage{multirow}
\usepackage{adjustbox}
\usepackage{tablefootnote}
\def\BibTeX{{\rm B\kern-.05em{\sc i\kern-.025em b}\kern-.08em
    T\kern-.1667em\lower.7ex\hbox{E}\kern-.125emX}}

\begin{document}

\title{A Distributed Inference System for Detecting Task-wise Single Trial Event-Related Potential in Stream of Satellite Images
{\thanks{This work was supported by the Agency For Defense Development Grant
Funded by the Korean Government (UI233001TD).}
}
}

\author{\IEEEauthorblockN{Sung-Jin Kim}
\IEEEauthorblockA{\textit{Dept. of Artificial Intelligence} \\
\textit{Korea University} \\
Seoul, Republic of Korea \\
s\_j\_kim@korea.ac.kr} \\
\IEEEauthorblockN{Dae-Hyeok Lee}
\IEEEauthorblockA{\textit{Dept. of Brain and Cognitive Engineering} \\
\textit{Korea University} \\
Seoul, Republic of Korea \\
lee\_dh@korea.ac.kr}
\and
\IEEEauthorblockN{Heon-Gyu Kwak}
\IEEEauthorblockA{\textit{Dept. of Artificial Intelligence} \\
\textit{Korea University} \\
Seoul, Republic of Korea \\
hg\_kwak@korea.ac.kr} \\
\IEEEauthorblockN{Ji-Hoon Jeong}
\IEEEauthorblockA{\textit{Dept. of Computer Science} \\
\textit{Chungbuk National University} \\
Cheongju, Republic of Korea \\
jh.jeong@chungbuk.ac.kr}
\and
\IEEEauthorblockN{Hyeon-Taek Han}
\IEEEauthorblockA{\textit{Dept. of Artificial Intelligence} \\
\textit{Korea University} \\
Seoul, Republic of Korea \\
ht\_han@korea.ac.kr} \\
\IEEEauthorblockN{Seong-Whan Lee}
\IEEEauthorblockA{\textit{Dept. of Artificial Intelligence} \\
\textit{Korea University} \\
Seoul, Republic of Korea \\
sw.lee@korea.ac.kr}
}

\maketitle

\begin{abstract}
Brain-computer interface (BCI) has garnered the significant attention for their potential in various applications, with event-related potential (ERP) performing a considerable role in BCI systems. This paper introduces a novel Distributed Inference System tailored for detecting task-wise single-trial ERPs in a stream of satellite images. Unlike traditional methodologies that employ a single model for target detection, our system utilizes multiple models, each optimized for specific tasks, ensuring enhanced performance across varying image transition times and target onset times. Our experiments, conducted on four participants, employed two paradigms: the Normal paradigm and an AI paradigm with bounding boxes. Results indicate that our proposed system outperforms the conventional methods in both paradigms, achieving the highest $F_{\beta}$ scores. Furthermore, including bounding boxes in the AI paradigm significantly improved target recognition. This study underscores the potential of our Distributed Inference System in advancing the field of ERP detection in satellite image streams.
\end{abstract}

\begin{IEEEkeywords}
event-related potential, target detection, deep learning, electroencephalogram, satellite images
\end{IEEEkeywords}

\section{INTRODUCTION}
Brain-computer interface (BCI), also known as brain-machine interfaces, represents a direct communication pathway between the brain and external devices. This technology has been at the forefront of neuroscience and engineering research, aiming to restore lost sensory and motor functions, enhance human capabilities, or even create new forms of interaction with the world. BCI decodes neural signals, translating them into commands that can control various devices or software applications \cite{lee2019towards, lee2020neural, thung2018conversion, lee2023autonomous}. They have been employed in medical settings to assist patients with severe motor disabilities, allowing them to communicate or control prosthetic limbs using only their brain activity \cite{kim2015abstract, lee2019possible}. In the realm of rehabilitation, BCI aids in restoring motor functions for stroke victims or individuals with spinal cord injuries \cite{nicolas2012brain, lee2019comparative}. Beyond medical applications, BCI has ventured into the entertainment industry, enabling users to play video games or operate virtual reality environments using their thoughts \cite{kerous2018eeg, bonnet2013two}. Furthermore, in the domain of defense and aerospace, BCI is being explored for piloting drones or even potentially controlling advanced machinery \cite{lee2021subject, kim2022eeggram}. The integration of event-related potential (ERP) within BCI, especially in tasks like satellite image analysis, underscores the potential of this technology in rapidly processing vast amounts of visual data, capturing the brain's instantaneous reactions, and making timely decisions.

In the realm of ERP detection, various methodologies have been proposed and explored over the years. The conventional approaches primarily relied on theory-driven methods, where the extraction of ERP components was based on predetermined temporal and spatial filters \cite{suk2014predicting}. These methods, such as independent component analysis and time-frequency representations, have been instrumental in isolating specific ERP components from the background EEG activity. Their strength lies in their ability to leverage prior knowledge about the expected ERP waveforms and their temporal dynamics.

However, with the advent of deep learning, the landscape of ERP detection has witnessed a paradigm shift. Deep learning-based methods, particularly convolutional neural networks and recurrent neural networks, have shown promise in automatically extracting intricate patterns from raw EEG data without the need for manual feature engineering \cite{mane2021fbcnet, bang2021spatio, kim2019subject}. These models are trained on enormous datasets, enabling them to capture the subtle nuances of ERP signals, even in noisy environments. Recent studies have demonstrated the superiority of deep learning techniques over the conventional methods, especially in scenarios with high inter-subject variability or when dealing with single-trial ERP detection \cite{zang2021deep}.

Detecting single-trial ERP presents a unique set of challenges. The inherent variability in neural responses, even to identical stimuli, could lead to the significant differences in ERP features across trials. This variability is further increased when considering the transition times between images. As the brain processes rapidly changing visual stimuli, the ERP elicited by one image can be influenced by the preceding image, leading to a temporal smearing effect. This effect complicates the task of isolating the ERP corresponding to a specific image, especially when the transition times between pictures are inconsistent.

In recent years, several methods have been proposed to address these challenges. Techniques such as adaptive filtering and advanced artifact rejection algorithms have been introduced to enhance the clarity of single-trial ERP \cite{woehrle2015adaptive}. Furthermore, deep learning architectures incorporating temporal attention mechanisms have been developed to focus on relevant time points in the ERP, mitigating the effects of varying transition times \cite{chen2022novel}. However, while these methods have shown promise, they often require extensive training data and may not generalize well across different subjects or ERP paradigms. The investigation for a robust and universally applicable single-trial ERP detection method remains an active area of research.

We have developed a novel approach, the Distributed Inference System, to address the inherent difficulties faced by the conventional single-trial ERP detection methods. Our proposed strategy diverges from the conventional approaches that attempt to process all diverse scenarios, such as varying image transition times and ERP stimuli, and tasks, like target recognition and target onset time inference, within a single model. Such an approach often restricts the model's ability to learn features from the data effectively. Our system adopts a strategy where each task is learned and inferred independently. By segmenting the learning process, each model in the system could learn about specific features relevant to each task, enhancing its overall accuracy and efficiency. This methodology has proven particularly effective in tasks involving the detection of targets within a stream of satellite images compared with the conventional methods.

\section{METHODS}

\subsection{Subjects and Experimental Environment}
We collected data from a total of four healthy subjects, comprising three males and one female. Our experiment received approval from the Institutional Review Board at Korea University [KUIRB-2022-0373-01]. Before the actual experiment, all participants were thoroughly briefed on the experimental paradigm and underwent practice sessions to familiarize themselves with the procedure. For EEG signal acquisition, we utilized a signal amplifier (BrainAmp, Brain Products GmbH, Germany). The EEG signals were sampled at a frequency of 250 Hz, and a 60 Hz notch filter was employed to eliminate DC noise. The EEG data was acquired using 32 channels positioned on the subjects' scalps following the international 10-20 system. To ensure optimal signal quality, we applied a conductive gel to the subjects' scalps, reducing the impedance of the EEG electrodes to 15 k$\Omega$ or below.

\begin{figure}[t!]
\centering
\scriptsize
\centerline{\includegraphics[height=0.38\textwidth]{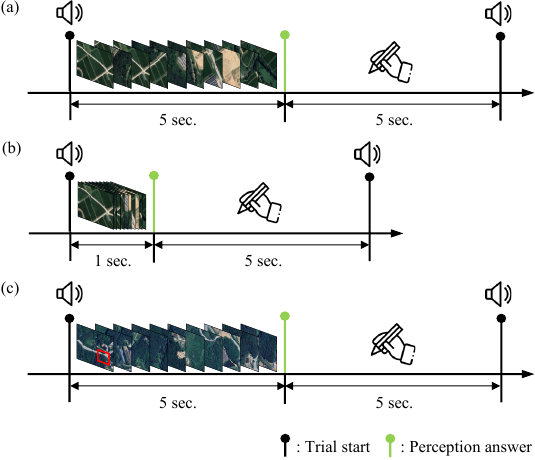}}
\caption{Experimental paradigm applied in the target recognition. (a) and (b) indicate the normal paradigm with the image transition time of 0.5 sec. and 0.1 sec., respectively. (c) indicates the AI paradigm containing a bounding box at images.}
\end{figure}

\begin{figure*}[t!]
\centering
\scriptsize
\centerline{\includegraphics[width=0.965\textwidth]{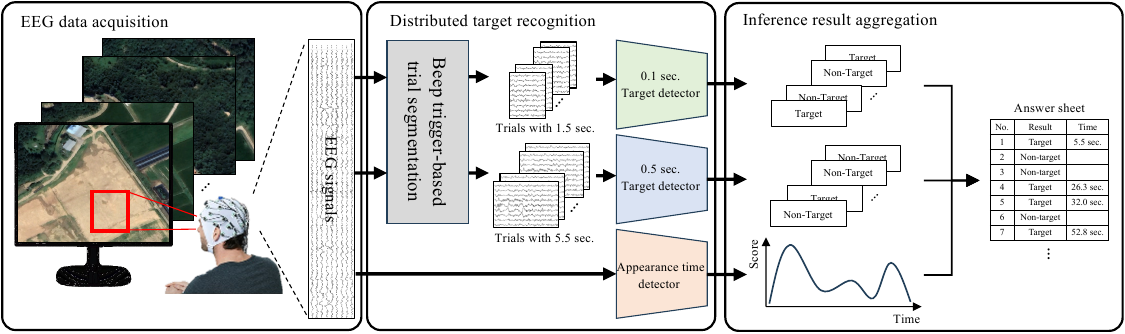}}
\caption{Overall flow to decide the decisions of target recognition using the proposed Distributed Inference System}
\end{figure*}

\subsection{Experimental Paradigm}
Our data acquisition paradigm centered on detecting targets within a sequence of 10 consecutively displayed satellite images. We conducted two distinct paradigms for this purpose. In the first paradigm (Normal paradigm), participants were presented with 16 trials, each comprising a series of satellite images. These trials were divided into two categories based on the transition time between images: 0.1 sec. and 0.5 sec. Of the 16 trials, 8 had a transition time of 0.1 sec., while the remaining 8 had a transition time of 0.5 sec. Targets appeared a total of 8 times across these 16 trials, with their appearance in the 0.1-sec. and 0.5-sec. trials being randomized. The second paradigm (AI paradigm) also consisted of 16 trials but with a consistent image transition time of 0.5 sec. across all trials. Unlike the first paradigm, this one simulated a scenario where the images had been processed by a computer vision model. As a result, targets within the images were marked with a red bounding box. However, considering the simulated computer vision inference, some non-target areas resembling the target were also mistakenly boxed in red. As with the first paradigm, targets appeared 8 times across the 16 trials. Each trial is separated by a beep sound that is inaudible to the subject in both paradigms. For both paradigms, one session consists of 16 trials, and four sessions are conducted in each paradigm. During both paradigms, participants were instructed to manually indicate the presence or absence of a target at the end of each trial in 5 sec., ensuring an interactive and attentive engagement with the presented images.

\subsection{Distributed Inference System}
Our primary goal through the paradigm was to determine the existence of a target within a given trial and, if present, to ascertain the exact time of its appearance. While the conventional methods employing a singular model for target detection share the overarching goal of capturing the characteristics of the ERP, our paradigm, which contains varying image transition times (0.1 sec. and 0.5 sec.) and the inference of target onset times, presents unique challenges. The differences in features that each task-specific model could learn make it difficult for a single model to optimize for all tasks simultaneously. To address this issue, we developed the Distributed Inference System to integrate the outputs of separated models tailored for each task.

Our system is structured around three main components: EEG data acquisition, Distributed target recognition, and Inference result aggregation. The EEG data acquisition component is responsible for collecting data from a participant over a session, which includes 16 trials. Following the previous part, the Distributed target recognition part takes over, segmenting the acquired EEG signals based on beep triggers, distinguishing between the 0.1-sec. and 0.5-sec. transition times. The data split by trial and the original data before the split are entered into the 0.1 sec. Target detector, 0.5 sec. Target detector, and Appearance time detector, respectively, to get the respective results. Each of these models independently processes the data and generates its inference. Finally, the Inference result aggregation component converges the outputs from the models. It determines the existence of a target in each trial using the results from each target detector model. In addition, it detects the exact onset time of the target with the Appearance time detector. By conducting all these processes, we could facilitate the concurrent extraction of results for all 16 trials within the session.

To include the P300 feature, which occurs when the target appears in the last image, each trial was split into one trial by including an additional 0.5 sec. of data after the 10-image stream. The 0.1 sec. Target detector and 0.5 sec. Target detector were trained on their respective segmented data. In contrast, the Appearance time detector utilized the entire EEG data, employing a sliding window approach with a window size of 1 sec. and a 0.5-sec. overlap for both training and inference. The optimized DeepConvNet \cite{schirrmeister2017deep} architecture was used for all detector models.

\subsection{Performance Evaluation}
Our evaluation was conducted in a subject-dependent environment. Each of the four sessions was designated as test data in turn, with the remaining three sessions serving as training data. This process was repeated four times, ensuring each session was used as test data once. The final performance was determined by averaging the results from these four evaluations. We conducted separate evaluations for the two paradigms. To focus on the recall of target detection, we employed the $F_{\beta}$ score as our evaluation metric, setting $\beta$ to 2. This choice emphasizes the importance of recall in our task. The formula for calculating $F_{\beta}$ score is shown below: 
\begin{equation} 
    {F}_{\beta} = (1+{\beta}^2)\frac{(\text{precision} \times \text{recall})}{({\beta}^2\cdot\text{precision}) + \text{recall}}
\end{equation}

Additionally, to evaluate the variability across subjects, we also computed the standard deviation of the $F_{\beta}$ scores for each participant. For the training process, each model was trained for 20 epochs with a learning rate of 0.001. AdamW optimizer \cite{loshchilov2018decoupled} was used for optimization with the weight decay of 0.01. In addition, the cosine learning rate scheduler was utilized for training stability.

\section{RESULTS AND DISCUSSION}
Our proposed method was evaluated against the conventional methods including ShallowConvNet \cite{kim2022rethinking}, DeepConvNet \cite{schirrmeister2017deep}, and EEGNet \cite{lawhern2018eegnet}. The evaluation method for the conventional methods was consistent with the conditions applied to the Appearance Time Detector in our system. The conventional methods assessed the performances by sliding a window of 1-sec. duration, moving in 0.5-sec. increments, and determining whether the segment contained a target. If a target was detected at least once within a trial by these methods, the entire trial was classified as containing a target. For both the traditional models and our Appearance Time Detector, a target score exceeding 0.5 was considered indicative of target presence.

Our proposed method, as shown in Table \uppercase\expandafter{\romannumeral1}, outperformed the other comparison methods in both paradigms, achieving the highest $F_{\beta}$ scores of 0.6875 and 0.7266 for the Normal and AI paradigms, respectively. We verified that all methodologies generally achieved better performance in the AI paradigm, where bounding boxes were present in the images, compared to the Normal paradigm. We attribute this performance enhancement in the AI paradigm to the presence of the red bounding boxes, which likely made the targets more conspicuous to the subjects, aiding in target detection. However, when examining the standard deviation value, a measure of variability, the EEGNet model recorded the lowest values of 0.0928 and 0.0881 for the two paradigms. Our proposed method achieved values of 0.1308 and 0.1481, indicating the more significant performance variance across subjects compared to other methods. The increased variance in our method's performance is attributed to notably enhanced results in specific subjects rather than a lack of stability. Furthermore, we won first place in the 1st BCI Target Recognition Technology Competition using our proposed method.

\begin{table}[t!]
\caption{Comparison of performances among the conventional methods across different paradigms}
\renewcommand{\arraystretch}{1.2}
\setlength{\arrayrulewidth}{0.13mm}
\small
\begin{tabular}{c|cc|cc}
\hline
\multirow{2}{*}{Method} & \multicolumn{2}{c|}{Normal paradigm} & \multicolumn{2}{c}{AI paradigm}   \\ \cline{2-5} 
                        & $F_{\beta}$            & std.             & $F_{\beta}$          & std.            \\ \hline
ShallowConvNet \cite{kim2022rethinking}          & 0.5314            & 0.1314           & 0.6481          & 0.1421          \\
DeepConvNet \cite{schirrmeister2017deep}            & 0.6328            & 0.1753           & 0.6899          & 0.1641          \\
EEGNet \cite{lawhern2018eegnet}                 & 0.6425            & \textbf{0.0928}  & 0.6841          & \textbf{0.0881} \\
\textbf{Proposed}       & \textbf{0.6875}   & 0.1308           & \textbf{0.7266} & 0.1481          \\ \hline
\end{tabular}
\footnotesize{{$^*$std.: standard deviation}} \\
\end{table}

\section{CONCLUSIONS AND FUTURE WORKS}
In this study, we introduced the Distributed Inference System, a novel approach designed to address the challenges of detecting single-trial ERP in a stream of satellite images. Our method, distinct from the conventional single-model approaches, employs multiple models tailored for specific tasks, ensuring optimized performance across varying image transition times and target onset times. The results demonstrate the superiority of our proposed system over existing methods in both the Normal and AI paradigms. Notably, the inclusion of bounding boxes in the AI paradigm enhanced target recognition, underscoring the potential benefits of visual aids in ERP detection tasks. However, while our system excels in post-session inference, it currently lacks the capability for real-time analysis. In future work, a primary objective will be developing a real-time inference system. Such an advancement would not only elevate the practicality of our approach but also pave the way for broader applications in real-world scenarios.

\bibliographystyle{IEEEtran}
\bibliography{REFERENCE}

\end{document}